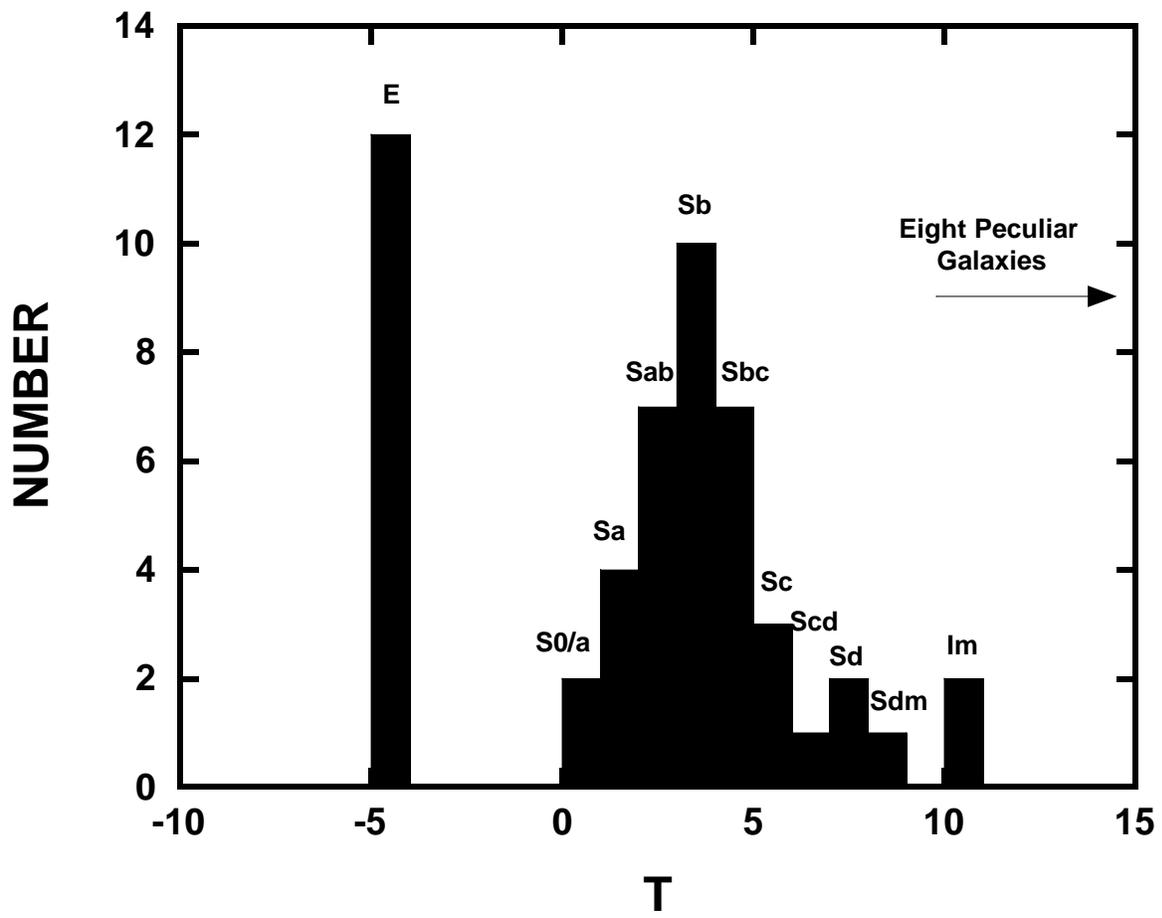

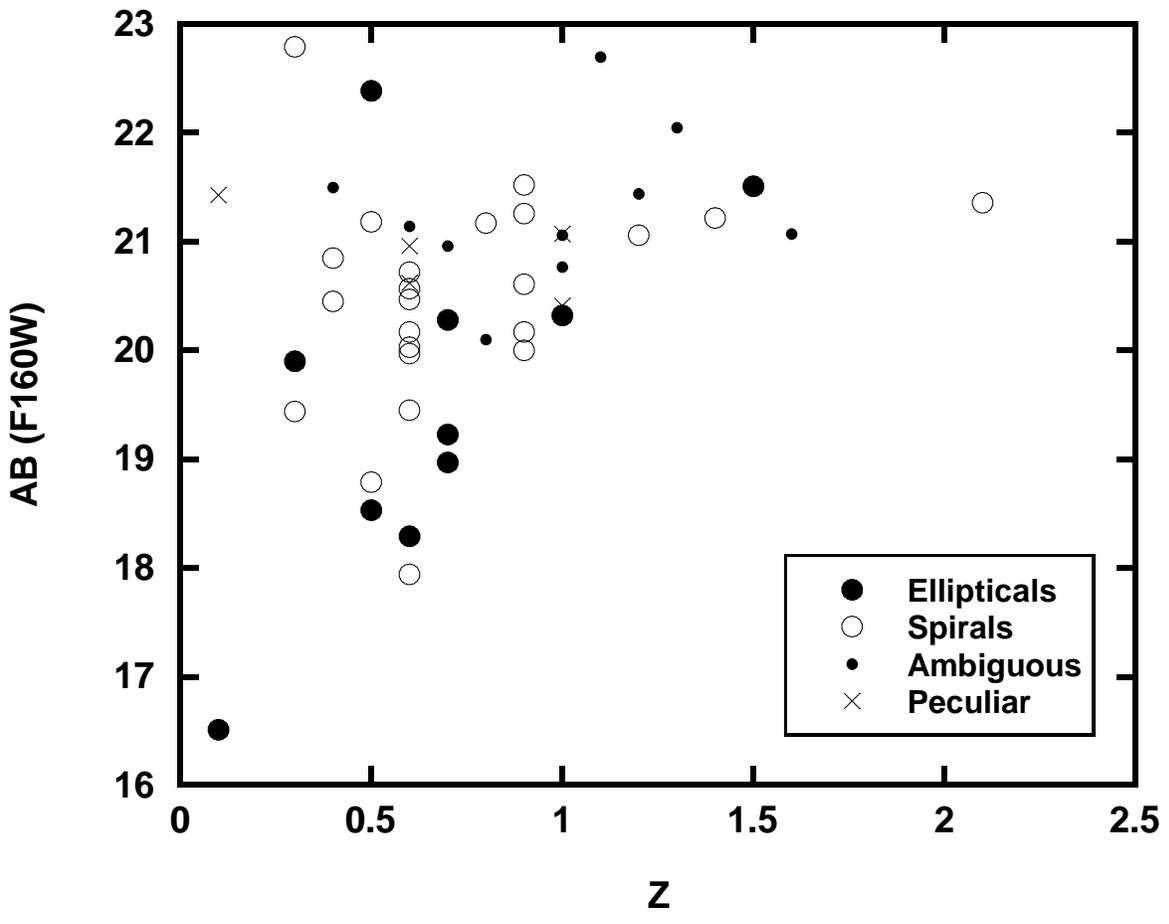

# Photometric Redshifts and Morphologies of Galaxies

# in the NICMOS Parallel Fields

Michael R. Corbin[1], William D. Vacca[2], Earl O'Neil[1],

Rodger I. Thompson[1], Marcia J. Rieke[1] and Glenn Schneider[1]

[1]NICMOS Group, Steward Observatory, The University of Arizona,

Tucson, AZ 85721; mcorbin@as.arizona.edu

[2]Institute for Astronomy, The University of Hawaii

2680 Woodlawn Dr., Honolulu, HI 96822; vacca@athena.ifa.hawaii.edu

ABSTRACT

We present positions, magnitudes, sizes and morphological classifications for 111 galaxies discovered in the *Hubble Space Telescope* NICMOS Camera 1 and Camera 2 parallel fields. We combine the magnitudes measured in the *JHK*-analog filters with those from deep ground-based images in *V* and/or *R* to measure photometric redshifts for 71 objects using Bruzual-Charlot population synthesis models. We find that these objects fall in the range *z* ~ 0.0 - 2.7, with <*z*> ≅ 0.8 and a mean luminosity <L> ≅ 1.6 L*. The NICMOS images reveal many of the galaxies to be ordered spirals and ellipticals similar to those in the local universe, with a high degree of symmetry and brightness profiles that are well-fit by de Vaucouleurs' $r^{1/4}$ and exponential disk laws. However, we find a higher fraction (~ 14%) of morphologically peculiar and/or interacting galaxies in the sample than is observed among local galaxies (~ 3% - 4%). This is consistent with the result from other deep HST images including the Hubble Deep Field and Hubble Medium Deep Survey field that the fraction of peculiar and interacting galaxies increases with redshift. As the NICMOS images of the sample galaxies cover their rest-frame near-infrared and optical emission, this result increases confidence that such changes in morphology are genuine, as opposed to an effect produced by viewing galaxies in the rest-frame ultraviolet.

We also find that at least 26 of the sample galaxies appear to be members of (non-interacting) pairs or groups, based on their proximity to one another and photometric redshifts. This is consistent with the results of recent ground-based optical surveys for faint galaxies covering larger areas, and with the detection of galaxy groups and filaments at redshifts higher than those covered by the present sample.

## 1. INTRODUCTION

During Cycle 7 of the *Hubble Space Telescope* (HST), the Near Infrared Camera and Multi-Object Spectrometer (NICMOS) instrument obtained several thousand images of random fields in parallel with pointed observations by the other HST instruments. The galaxies in these parallel images have been the subject of several studies (Teplitz et al. 1998; Yan et al. 1998; Yan et al. 1999; McCarthy et al. 1999; Treu & Stiavelli 1999), primarily for the determination of their number counts. However, with the exception of 33 relatively bright emission-line galaxies identified in the grism images obtained by NICMOS Camera 3 by McCarthy et al. (1999), no redshifts have been measured for these objects. Redshifts of these galaxies can be used for the study of galaxy morphological evolution, and will also aid the effort to probe galactic dynamics and star formation beyond the local universe by establishing targets for more intensive observations. Importantly, by covering the rest-frame optical emission of galaxies at $1 < z < 3$, near-infrared images provide a "morphological *K*-correction" that is necessary to allow the comparison of the morphologies of these objects with the optical morphologies of nearby galaxies. Comparison of the optical and near-infrared morphologies of intermediate-redshift galaxies in the Hubble Deep Field (HDF) has shown morphological *K*-corrections to be important, insofar as several intermediate-redshift galaxies with fragmentary appearances in the rest-frame ultraviolet are found to have smooth underlying light distributions in the rest-frame optical (Bunker, Spinrad & Thompson 1999). The discovery of these galaxies in the near infrared also avoids selection effects associated with the optical detection of galaxies in this redshift range, in particular, high star formation rates. This is important for measuring the evolution of galactic star formation rates from early epochs to the present.

In this paper we present data on 111 galaxies discovered in the NICMOS parallel images obtained with the high-resolution Cameras 1 and 2. Combining these data with ground-based optical images, we are able to measure photometric redshifts for 71 of the sample objects. We have additionally undertaken a more detailed analysis of the object morphologies than was presented previously (Teplitz et al. 1998), and compare these results with our redshift estimates. Our results should be particularly useful as a basis for follow-up spectroscopy of the objects with very large telescopes and for observations at other wavelengths.

In the following section we discuss our data and object selection criteria. In § 3 we present our results, and conclude with a discussion in § 4. $H_0 = 50$ km s$^{-1}$ Mpc$^{-1}$ and $q_0 = 0.1$ are assumed throughout.

2. DATA

2.1 The NICMOS Parallel Fields

Parallel images in NICMOS Cameras 1 and 2 were obtained during the first part of HST Cycle 7, from 1997 June 2 to 1997 Nov 8. Parallel imaging was then switched to Camera 3 as it came into better focus (see Thompson et al. 1998), due to the larger field of view (51.2″ × 51.2″ versus 11″ × 11″ and 19.2″ × 19.2″ for Cameras 1 and 2, respectively) and the availability of its grisms. However, the lower spatial resolution of Camera 3 (0.2″ pix$^{-1}$, versus 0.043″ pix$^{-1}$ for Camera 1 and 0.075″ pix$^{-1}$ for Camera 2) offers only marginal resolution of intermediate- and high-redshift galaxies, which WFPC2 imaging has revealed to be ~ 1″ in size (e.g. Giavalisco, Steidel & Macchetto 1996; Steidel et al. 1996). We have thus restricted our analysis to the Camera 1 and Camera 2 images. The Camera 1 parallel images at a given spacecraft pointing were obtained in the F110W and F160W filters, which have central wavelengths and FWHM values of 1.102 μm, 0.592 μm and 1.593 μm, 0.403 μm, respectively. The Camera 2 parallel images were obtained in the same filters, and also in the F222M filter, which has a central wavelength of 2.216 μm and a FWHM of 0.143 μm. We tried to obtain optimum quality of the images by reducing them separately from the "pipeline" reduction procedure applied to the images placed in the HST public archive. To do this, we applied a set of averaged dark current and flat-field images obtained on-orbit and close in time to the parallel images themselves with the IRAF.STSDAS task CALNICA (Bushouse 1997). By contrast, the pipeline reductions use flats obtained during NICMOS laboratory testing, and a set of synthetic dark current images. The parallel images were taken in the MULTIACCUM mode, in which the detectors are non-destructively read out several times in the course of completing the total integration. In principle this removes the cosmic rays from the images, but in practice additional cosmic ray removal is necessary, and for the parallel images was performed with the IRAF task "cosmicrays." We also replaced bad pixels on the detectors using averaged values of surrounding pixels. We estimate an improvement of 3% - 5% in the relative and absolute photometric accuracy of our images compared to those available in the archive. Our final set of images covers 342 separate fields over a wide range of Galactic latitudes, with a total area coverage of approximately 8.7 arcmin$^2$ for Camera 1, and 26.6 arcmin$^2$ for Camera 2. This is approximately five times the area covered by the WFPC2 observations of the HDF.

Integration times of the parallel images (the same for each camera) vary between 192s and 1280s per filter. Significantly higher effective exposure times were obtained after averaging together multiple images of a given field, as well as those images having significant spatial overlap. The sensitivity level of the final set of combined images is therefore very non-uniform, and so it is hard to assess completeness and interpret source counts (but see Yan et al. 1998 and Teplitz et al. 1998). In Table 1 we present the coordinates of the centers of all 75 of the fields in which one or more galaxies were detected, along with the camera in which they were detected and the 1$\sigma$ value of the background level of the final image in each filter in micro-Janskys. The mean sensitivity levels for each filter are given at the end of the table.

## 2.2 Object Selection Criteria

Galaxies in the set of re-reduced Camera 1 and Camera 2 parallel images were initially selected by eye, under the criteria that the object display a clearly extended structure, is not blended with another object, and does not fall on or near the edge of the image. Our final criterion, applied in the most uncertain cases, is that the object have an integrated flux ~ 3$\sigma$ above the local background level, although we do not attempt a morphological classification for objects with integrated fluxes less than ~ 10$\sigma$ above the background level. Numerous other objects marginally miss the detection threshold, and would likely meet our selection criteria in deeper images; the sample is thus likely biased towards more luminous galaxies. Our final sample consists of 11 objects detected by Camera 1 and 100 objects detected by Camera 2.

Figure 1 shows the F160W band images of 16 objects from the total sample of 111. We have organized the objects by morphological type, discussed in greater detail in § 3, including spirals, ellipticals, objects with ambiguous morphology, and objects that are peculiar, and possibly interacting or merging. The contrast levels of the images have been adjusted to best show the galaxy morphologies, e.g., to reveal both the bulge and disk among the objects that appear to be spirals. We have labeled the objects using their J2000 coordinates as measured from the image header files, with the prefix NPF indicating their discovery in the NICMOS parallel fields. The accuracy of the object coordinates is limited by the pointing accuracy of HST itself, which is ~ ± 0.5″.

2.3 Optical Observations

We obtained CCD images in the Harris *V* and/or *R* bands of 62 of the sample objects, using the 61" Kuiper reflector and 90" Bok reflector of Steward Observatory. Integration times varied between 45 minutes and 2 hours, which should provide sensitivities comparable to the parallel images. Seeing was typically mediocre (~ 1″ - 2″), which reduces this sensitivity, particularly for objects of low surface brightness, whose flux may be spread out below the detection level. Conditions were photometric for approximately two-thirds of the target fields, and stars in the Landolt (1992) fields were used for calibration. For observations made under non-photometric conditions, the images were multiplicatively scaled to match those taken under photometric conditions. Astrometric registration to the NICMOS images was obtained by aligning the optical images with Digital Sky Survey images centered at the position of the NICMOS images, which is important in cases in which the NICMOS image contains no bright stars or objects other than the candidate galaxy. This procedure in principle provides astrometric alignment of the images to within 1″. Nonetheless, in two cases the correspondence of objects in the optical and infrared images is uncertain, and the associated optical magnitude limits for these objects have been marked with a question mark in Table 2, which presents the objects' optical and infrared magnitudes. As Table 2 also shows, in many cases we failed to detect the object in either the *V* or *R* images, and we are only able to place lower limits on its magnitudes in these bands.

A comparison of the *R*, F110W, F160W, and F222M images of a Camera 2 field containing two galaxies from the sample is shown in Figure 2. Although they have different spectral energy distributions, our redshift estimates of these two objects (discussed in the following section) are equal to within uncertainties, so they may be physically associated.

3. ANALYSIS

3.1 Photometry

Aperture photometry of the objects was obtained for both the NICMOS and optical images using the IRAF task "polyphot," which allows polygonal apertures to be custom fit around each object, with the aperture boundaries set at the points where the object flux fell below the sky level. The sky level was measured from a circular annulus with an inner radius set ~ 1″ - 2″ from the aperture boundary of each object and placed to avoid including any nearby objects. Magnitudes measured in this way should be more accurate than those measured from an aperture fixed in size and shape, which may include extraneous objects. The resulting instrumental magnitudes were converted to fluxes from the standard star observations noted in § 2.3 for the optical observations, and from the standard star observations made on-orbit for the NICMOS cameras (see the NICMOS section of the Space Telescope Science Institute website, www.stsci.edu/instruments/nicmos/, and Colina & Rieke 1997). Corrections for Galactic extinction were applied using the reddening maps of Schlegel, Finkbeiner & Davis (1998); for the NICMOS data, we used an interpolation of the reddening law derived by Rieke & Lebofsky (1985) to convert from E(B-V) to infrared extinctions. Uncertainties in these calibrations and within the data themselves are relatively large, and we estimate accuracy in the resulting magnitudes and fluxes to be only ~ 10% - 20%. Several objects within the sample appear to have companions (see Figure 1*c*), but the fainter object in most cases cannot be reliably measured.

Magnitudes for each object in each band are given in Table 2. Magnitudes are given in the AB system, determined from the relation M = -2.5log$f_\nu$ - 48.58, where $f_\nu$ is the object flux in ergs

cm$^{-2}$ s$^{-1}$ Hz$^{-1}$. Limits on these magnitudes for the objects not detected in a given band are also given, although these are not particularly useful in the case of the F222M observations, due to the relatively low sensitivity associated with this narrower bandpass and the higher level of the background emission at the filter wavelength.

3.2 Sizes, Morphologies and Brightness Profiles

Table 2 lists the first moment radius, in arcseconds, of each object as measured in the F160W images. This radius was also used for the objects in the HDF by Williams et al. (1996) and is defined as

$$r_1 \equiv \Sigma rI(r) / \Sigma I(r)$$

where $I(r)$ is the object intensity as a function of distance measured from the peak value to a level 3$\sigma$ above the background. For several objects the value of $r_1$ is equal to or near the FWHM of the image point-spread function, indicating that the light from the galaxy is dominated by an unresolved core, most likely an AGN or nuclear starburst. This is consistent with our visual inspection of the images, which reveals many of the objects to have bright cores, and also with the results of McCarthy et al. (1999), who find many of the galaxies discovered in the NICMOS Camera 3 grism images to be actively star-forming.

In Table 2 we also provide two estimates of each object's morphology. First, we note whether the galaxy appears to be an elliptical, spiral, peculiar or whether its morphology is ambiguous. Second, for the objects whose morphology is not ambiguous, we estimate the morphological "T" parameter used in the Third Reference Catalog of Bright Galaxies (de Vaucouleurs et al. 1991) and subsequent studies of galaxies in the local universe (e.g. Marzke et al. 1994). This parameter has range of -6 to 0 for elliptical galaxies, 0 to 9 for spirals (increasing from earlier to later types), a value of 10 for irregular galaxies, and a value of 99 for peculiar galaxies. We estimate the uncertainty in the value of this parameter that we have assigned to be typically ± 1. In particular, the faintness of the objects makes it difficult to distinguish between S0, Sa and elliptical galaxies, and some of the ellipticals may have T values in the range -3 to -1, although we have assigned them all values of -5. The distribution of the T values is shown in Figure 3.

Our morphological classifications are based both on the appearance of the galaxies in all bands and on attempts to fit their brightness profiles using the IRAF.STSDAS tasks "ellipse" and "nfit1d." The latter task allows the fit of a deVaucouleurs' $r^{1/4}$ and exponential disk brightness profile of the forms

$$I(r) = I_0 \exp(-7.688[(r/r_0)^{1/4} - 1])$$

$$I(r) = I_0 \exp(-r/r_0)$$

Although galaxy profile fitting is best performed after deconvolution with the image point-spread functions, we did not attempt this for two reasons. First, most of the images containing galaxies did not have a star or other point source from which to determine the PSF, and measurement of the PSF in other Camera 1 and Camera 2 parallel images revealed a relatively wide range (exceeding typical measurement errors) in the value of the PSF FWHM that was likely produced by focus variations in the cameras and spacecraft aberration over the period of the observations. A single "mean" PSF could thus not be appropriately applied to the images for deconvolution. Second, for objects not obviously spirals or ellipticals, the uncertainty arises mainly from a limitation in the signal-to-noise level in the object, rather than a limitation in resolution (e.g. Fig. 1$c$).

For approximately 55% of the sample objects, either the "ellipse" or "nfit1d" tasks failed to converge on a solution. However, for most of the objects for which such failure occurred, the object's appearance did not clearly rule out the possibility of its being a spiral or elliptical, and the failure of the fitting tasks is likely due to the low signal-to-noise ratio ($< 10$) in the detected emission. We have therefore classified such objects as having an ambiguous morphology, and show four examples in Figure 1$c$. By contrast, other objects for which the profile-fitting tasks failed have morphologies that are clearly not those of normal spiral and elliptical galaxies, and are distinguished by asymmetric brightness distributions and/or evidence of gravitational interaction with a close companion, or of being a galaxy merger in progress. We have classified such objects as being peculiar, noting cases where this peculiarity appears to be the result of a merger or interaction. The four most notable examples we find are shown in Figure 1$d$. Two other

objects, NPF 122758.60+170250.6 and NPF 125137.13+563016.4, have been classified as irregulars, as their appearance and brightness profiles are similar to local irregular galaxies, and they are likely to be at low redshift. Classification of an object as an elliptical was made only if an $r^{1/4}$ law provided a good fit to the brightness profile, and a disk was clearly not discernible. Classification of an object as a spiral was made only if its brightness profile could be fit by the combination of a bulge and disk law. Further discussion of the properties of the individual galaxies is given in the Appendix. Among the galaxies that we are able to classify, the numbers of spirals, ellipticals, peculiars and irregulars are 37:12:8:2.

Figure 4 shows the azimuthally averaged brightness profiles, along with our fits to them, for four objects from the sample that we have classified as spirals and ellipticals. The F160W images of these objects are shown in Figures 1*a* and 1*b*. A combined bulge and disk profile provides a very good fit to the objects NPF 122751.33+170229.0 and NPF 153215.95+324743.8, while the objects NPF 033321.07-092236.2 and NPF 223000.78+265128.5 are well fit by an $r^{1/4}$ law alone, with no evidence of a disk, and have been classified as ellipticals. The brightness profiles of the two spiral galaxies, along with those of some of the other galaxies in the sample classified as spirals, show evidence of the type of "lens" structure at the bulge/disk junction found among some nearby spirals (Kormendy 1982 and references therein). We note that we are able to obtain the same quality of fit to the profiles of these objects both with and without truncating the inner disk radius at the apparent position of the bulge. We also note that we find no clear evidence of barred spirals, in contrast to what has been found from the NICMOS F110W and F160W images of galaxies in the Hubble Deep Field (Bunker, Spinrad & Thompson 1999). As may be seen in Figure 1*a*, the disks of the spiral galaxies show no strong evidence of spiral arms, indicating that the detected emission is dominated by their Population II stars, as would be expected for near-infrared observations of galaxies at z < 3.

Table 3 lists the ratio of major to minor axis length among the spiral galaxies in the sample, along with the position angles of their major axes. This latter quantity is relevant to follow-up spectroscopy, particularly for the highly inclined objects.

### 3.3 Spectral Energy Distributions

### and Photometric Redshifts

Bruzual & Charlot (1993) investigate galactic spectral evolution using stellar population synthesis models, assuming several different modes of star formation. We use the 1995 versions of their model spectra to study the spectral energy distributions (SEDs) of our sample objects, and to estimate their photometric redshifts. The object SEDs, as inferred from the measured magnitudes, suggest a fairly wide range of ages and star formation rates. A study by Sawicki & Yee (1998) of galaxies in the HDF selected by the Lyman break method has also indicated that their star formation proceeds episodically rather than continuously. Assuming that the star formation of our sample objects is also proceeding episodically, we thus chose the Bruzual & Charlot model for the case of an instantaneous burst of star formation. This can serve as an approximation to a situation in which our sample galaxies are at various points in a cycle of short star-forming episodes followed by periods of quiescence.

We selected the Bruzual & Charlot models for the case of solar metallicity and a Salpeter initial mass function with a range of 0.1 $M_\odot$ to 125 $M_\odot$. Spectra for these input parameters at ages of 0.001, 0.01, 0.1, 0.3, 0.6, 1, 4, 7, 13 and 19 Gyr were generated and then used as templates to fit the object SEDs over the range $z = 0 - 6$, using a $\chi^2$ minimization procedure. The redshifts associated with the lowest values of $\chi^2$ are given in Table 2, for the objects with fluxes in three or more bands. Uncertainties in these values are typically $\pm$ 0.1. In some cases the distribution of the $\chi^2$ values for the spectral templates had two minima, corresponding to two sets of ages and redshifts which provided a good fit to the observed SED. This is due to an ambiguity between the Lyman and Balmer continuum breaks in the generated spectra. To resolve this ambiguity, we measured the ratio of the object rest-frame *V* band luminosity to L* (using $M_V^* = -21.5$; Mihalas & Binney 1981). In nearly all cases where a value of $z > 3$ was indicated by the template fit, the associated luminosity was implausibly large (> 10 L*), and so for these objects we adopt the lower redshift value. For the adopted redshifts, we find $4 \times 10^{-2} < L/L^* < 4.7$, with a mean value of approximately 1.6 L*, consistent with the expected luminosity bias of the sample (§ 2.2). The redshifts of objects with values of $\chi^2$ exceeding ~ 3, indicating a poor fit, as well as cases where a large range of redshift produced an acceptable fit, have been marked with a colon in Table 1. For every object with an acceptable fit, the associated template was in the 0.1 - 7 Gyr age range, and we find that the associated ages decrease approximately with redshift. These ages are, strictly speaking, the age since the last star formation episode, and not that of the entire galaxy, and so may be considered lower limits. Four examples of the template fits to the object SEDs are shown in Figure 5.

The Bruzual & Charlot models do not include the effects of internal dust, emission lines, or absorption by the intergalactic medium, all of which will affect the SEDs of intermediate- and high-redshift galaxies at some level (see, e.g. Corbin et al. 1998; Sawicki & Yee 1998; Armus et al. 1998). Some of the galaxies are also likely to host AGNs (see the Appendix). However, including these parameters in our template fits would be inappropriate, given the small number of flux points available. Additionally, none of the objects appear to be above $z \sim 4$, the point at which the Lyman break region enters the *V* band, and thus where extinction by internal dust and absorption by the IGM have the largest effect on the observed fluxes in our passbands. For objects at lower redshift and for those without *V* band observations, the portion of the SED measured (the rest-frame near infrared and optical) will not be as strongly affected by internal extinction and IGM absorption as the rest-frame ultraviolet. Fernandez-Soto, Lanzetta & Yahil (1999) have also found a good correlation between photometric and spectroscopic redshifts for galaxies in the HDF, based on combined optical and near-infrared images for the former, which increases our confidence in these photometric redshift values.

In Figure 6 we plot F160W AB magnitude versus photometric redshift, excluding objects having large uncertainty in the latter. As expected, the number of objects with ambiguous morphologies increases at larger magnitudes and redshifts. There are no clear differences between the magnitude and redshift distributions of the other morphological types. We find mean redshift values for each morphological type of $<z> = 0.75$ (S), $<z> = 0.66$ (E), $<z> = 0.97$ (A), and $<z> = 0.66$ (P), with an overall sample mean of 0.78. However, we do note that among the galaxies that we are able to morphologically classify, the percentage of peculiar galaxies (14%) is much higher than the percentage of galaxies in the local universe noted as having special, irregular or unclassifiable morphologies. Specifically, the percentages of such galaxies in the revised Shapley-Aames catalog (Sandage & Tammann 1981) and the Center for Astrophysics redshift surveys (Marzke et al. 1994) are ~ 3% - 4%, using samples in excess of 1,000 objects. With our criteria for assigning a morphological class to our objects (well-resolved with a signal - to-noise ratio > 10), this difference is not likely to be artificial, e.g. as a result of the $(1 + z)^4$ surface brightness dimming of normal spirals. For example, the objects NPF 034247.20-295043.8 and NPF 154927.20+211848.2 (Fig. 1*c*) are very unlikely to be misclassified spirals.

### 3.4 Evidence of Galaxy Pairs and Groups

A number of our sample galaxies are found within the same field, and have similar magnitudes and redshifts (e.g. Figure 2). Inspection of Table 2 reveals eight candidate galaxy pairs, whose photometric redshifts are equal to within the typical (~ ± 0.1) uncertainties in the population synthesis model fits, and whose physical separations assuming equal redshifts are at most approximately 125 kpc. We provide a summary list of these candidate pairs and their separations in Table 4. These object are distinguished from those that we have identified as mergers or interacting systems, insofar as such interaction is not obvious and the galaxy separations are larger. It would appear instead that these objects are members of the same groups, although the small field sizes and object faintness limit the ability to confirm other group members. However, in two Camera 2 fields we find candidate galaxy groups containing four members each, which are shown in Figure 7. The faintness of most of the objects in these fields has prevented us from measuring photometric redshifts for them, and thereby testing the group hypothesis. However, we can compare the apparent overdensity of objects in these fields to the galaxy counts in the NICMOS parallel fields measured by Teplitz et al. (1998) and Yan et al. (1998). They find that the F160W band surface density of galaxies in the AB magnitude interval 22 - 23, comparable to the range occupied by our candidate groups (Table 2), is $\log N \cong 4.64$ mag$^{-1}$ deg$^{-2}$. Our candidate groups, with lower limits of membership of four, exceed this surface density by a factor of approximately five. Similarly, the probability of eight galaxy pairs randomly arising within a sample of this size is < 0.1%. This strongly suggests that many of these objects are indeed members of groups, if not larger clusters.

## 4. DISCUSSION

The interpretation of our results must be tempered by the small number of galaxies in the sample that we are able to morphologically classify (59), and the associated bias towards more luminous objects. However, while the distribution of morphological types (Figure 3) is basically similar to that found among galaxies in the local universe (e.g. Sandage & Tammann 1981; Marzke et al. 1994), as noted in § 3.3 the percentage of peculiar galaxies is higher by a factor ~ 3 - 4. This is consistent with the results from the Hubble Deep Field, Hubble Medium Deep Survey Field and other HST deep imaging studies that the fraction of galaxies with peculiar morphologies and/or that are interacting increases strongly at intermediate redshifts. Specifically, studies including Burkey et al. (1994), Glazebrook et al. (1995) Abraham et al. (1996a, 1996b), Driver et al. (1998), Im et al. (1999) and van Dokkum et al. (1999) find a significantly higher percentage of peculiar/interacting galaxies out to $z \sim 1$ in these fields from the longest-wavelength (*I* band) observations made with the WFPC2. By covering the near-infrared and red portions of the continua of the objects in the present sample, the NICMOS observations increase confidence that these morphological differences are genuine and are not the result of the lack of a morphological *K*-correction.

The change in the proportion of spiral and peculiar galaxies with redshift in the Hubble Deep Field and Hubble Medium Deep Survey Field led Driver et al. (1998) and Im et al. (1999) to suggest that the latter are evolving into the former, following dynamical relaxation and the completion of the merging of sub-components. The present data cannot really address this issue, due to the small number of peculiar galaxies in the total sample, and the result that several of these peculiar galaxies, including NPF 034247.20-295043.8 and NPF 154927.20+211848.2 (Fig. 1*c*), appear to involve interaction between ellipticals. However, several of the peculiar and ambiguous galaxies in our sample, including NPF 065228.62-001338.0 (Figure 1*c*), are in double systems that appear very similar to recent simulated HST images of disk galaxies in the process of formation (Contardo, Steinmetz & Fritze-von Alvensleben 1998).

This motivates deeper HST imaging of these objects in more bands to study their morphologies in greater detail. Concerning the elliptical galaxies in the sample, we note that the spectral template fits indicate that they are among the oldest within it. For example, we find an age ~ 7 Gyr for the elliptical NPF 211655.38+023333.2, which at the associated photometric redshift of $z = 1.5$ points to a formation redshift above five. An object this old at such a redshift would also have important cosmological implications (see Stockton, Kellogg & Ridgway 1995; Dunlop et al. 1996), motivating spectroscopic confirmation of this photometric redshift and a more careful study of the object's spectral energy distribution, particularly to check for the presence of dust. Benítez et al. (1999) also find evolved elliptical galaxies at $z \sim 1.5$ in the Hubble Deep Field South, and Treu & Stiavelli (1999) argue for high formation redshifts ($z > 3$) for ellipticals based on the analysis of candidates discovered in the NICMOS Camera 3 parallel fields. In conjunction with our results, these discoveries run somewhat contrary to the findings of Zepf (1997) and Barger et al. (1999), which suggest lower formation redshifts for the elliptical population in general. Further work in this area is clearly needed. In particular, the nature of the population of "Extremely Red Objects" (see Thompson et al. 1999 and references therein), whether very dusty young galaxies or very old ellipticals at $z \sim 2$, remains unclear.

The evidence of galaxy pairs and groups in the sample discussed in § 3.4 is consistent with the results of recent deep ground-based surveys for faint galaxies covering much larger areas (De Mello et al. 1997a, 1997b; Small et al. 1999). Specifically, the De Mello et al. survey finds a comparable incidence of pairs and groups of galaxies within a total sample of over 84,000 objects, while the redshift survey of Small et al. finds evidence that the majority of galaxies out to $z \cong 0.5$ are members of larger structures. Our results are also consistent with the discovery of a galaxy group at a redshift of 2.38 (Francis et al. 1996; Francis, Woodgate & Danks 1997) based on the occurrence of paired absorption lines in two closely spaced QSOs. The $z = 2.38$ group has been argued to be part of a large filament of galaxies, which in turn would be consistent with the evidence of such structures among Lyman break galaxies whose redshifts have been spectroscopically confirmed (Steidel et al. 1998). The full extent of our candidate groups is poorly constrained by the small size of the Camera 1 and Camera 2 fields, and so it is not possible to assess whether they are only parts of clusters or larger structures. However, the available results indicate that at least galaxy pairs and groups were well in place by $z \sim 1$. Given the cosmological importance of the existence of galaxy clusters and filaments at intermediate and high redshift (e.g. Bahcall & Fan 1998), deeper imaging of larger fields centered on the candidate groups is needed.

Finally, we re-iterate that the galaxies discovered in the NICMOS parallel fields present the opportunity to investigate many more important questions. For example, measurement of the H$\alpha$ line in the galaxies at $z \sim 0.5 - 2$ can aid efforts to determine the evolution of star formation with cosmic time (see Madau, Pozzetti & Dickinson 1998; McCarthy et al. 1999), as there is a relative paucity of galaxies known at such redshifts due to the limitations of optical selection techniques. The edge-on spiral galaxies in the sample offer the chance to probe the rotational kinematics of such galaxies at early epochs, using H$\alpha$ and other nebular emission lines observable in the near infrared (see Pettini et al. 1998), and thereby examine the Tully-Fisher relation at higher redshifts. Likewise, the elliptical galaxies in the sample provide a means to investigate the evolution of the fundamental plane relations. Similarly, the identification of more galaxies at $z \sim 1$ can aid efforts to identify supernovae at intermediate redshift for the purpose of constraining cosmological parameters (Perlmutter et al. 1998; Garnavich et al. 1998). Deeper imaging with NICMOS and WFPC2 of the peculiar galaxies in the sample will also allow a more detailed comparison of their properties to the results of numerical simulations of galaxy formation, in order to better test the interpretation that they are the progenitors of more normal spirals and ellipticals.

We thank Betty Stobie, Howard Bushouse, and Ivo Busko for their assistance with the analysis software, Heidi Olson for her work on the images, and Teresea Brainerd, Nelson Cadwell, Roelof deJong, Richard Green, Huan Lin, Matthias Steinmetz and Ray Weymann for helpful discussions of the results. We also thank an anonymous referee for comments and suggestions that improved the paper. This work was supported by NASA grant NAG 5-3042 to the University of Arizona.

APPENDIX

COMMENTS ON INDIVIDUAL OBJECTS

NPF 002615.22+104856.8  --  Very bright core, possible AGN.

NPF 024249.29-340131.1  --  Two other marginal objects nearby; group?

NPF 024757.82+194626.3  --  Highly inclined spiral.

NPF 024757.55+194643.8  --  This appears to be a double system with a small, fainter member.

NPF 034247.20-295043.8  --  This object has a peanut-shaped morphology and a bright core, suggesting a merger and the presence of an AGN.  No evidence of a disk is seen, suggesting a merger between elliptical galaxies.

NPF 041229.42-573633.2  --  Face-on, early-type spiral.

NPF 053510.72+220308.2  --  Highly inclined spiral.

NPF 065228.62-001338.0  --  Apparent double with a small, interacting companion.

NPF 074351.72+651658.6  --  This spiral may be interacting with a smaller companion on its northeast side.

NPF 102427.40+465933.2  --  Low surface brightness or very late-type spiral viewed nearly face-on.

NPF 114929.42+124716.5  --  Low surface brightness.

NPF 125624.12+220452.7  --  Highly inclined spiral with an asymmetry between the opposite sides of the bulge; effect of a dust lane?

NPF 142315.18+383225.9  --  Very early-type spiral; S0?

NPF 154927.20+211848.2  --  This object appears to be a merger in a late stage, between an elliptical and a smaller companion.

NPF 155116.54+325655.9  --  Possible companion.

NPF 160936.82+652500.3  --  This spiral has a very bright core, suggesting the presence of an AGN.

NPF 160937.81+652456.5  --  Diffuse, low surface brightness spiral.

NPF 16320592-130736.3  --  This appears to be an interacting system similar to M 51, with the suggestion of a bridge of material between the objects.  Measured quantities are for the larger of the two objects, as the smaller object's could not be reliably measured.

NPF 211655.39+023332.0 -- This object may form a group with NPF 211655.43+023329.5 and

NPF 211655.54+023326.8.

NPF 221748.12+003138.1 -- This object has an unusually high F222M band flux, possibly by a coincident cosmic ray, and no spectral template could be fit.

FIGURE CAPTIONS

FIG. 1. -- NICMOS Camera 1 and Camera 2 F160W images of 16 objects in the present sample. Images are 3″ square and the orientations are random. The images show (*a*) spiral galaxies, (*b*) elliptical galaxies, (*c*) galaxies with ambiguous morphologies, and (*d*) galaxies with peculiar morphologies that appear to be interacting or merging.

FIG. 2. -- Comparison of the *R*, F110W, F160W and F222M images of the field containing the sample objects NPF 005810.72+302732.7 (top) and NPF 005810.36+302729.3. Field size is approximately 8″ square. The spot at the left edge of the F222M image is the Camera 2 coronagraphic hole. Our spectral template fits yield redshifts of 0.4 ($\pm$ 0.1) and 0.3 ($\pm$ 0.1) for these objects, respectively.

FIG. 3. -- Distribution of the T values of the sample galaxies, with the associated Hubble type noted. The eight peculiar galaxies in the sample have T values of 99.

FIG. 4. -- Azimuthally-averaged brightness profiles of four objects in the sample. The F160W images of the objects are shown in Figures 1*a* and 1*b*. Uncertainties in the flux values range from ~ $\pm$ 0.01 to $\pm$ 0.1 $\mu$Jy. The dashed lines represent the best-fitting de Vaucouleurs' $r^{1/4}$ law and exponential disk profiles, and the dotted line represents the combination of the two.

FIG. 5. -- Spectral energy distributions of four objects in the sample, shown in the observed frame, along with the best-fitting Bruzual-Charlot population synthesis model spectrum. The model age is given by $\tau$.

FIG. 6. -- Comparison of the object AB (F160W) magnitudes with photometric redshift, excluding objects with highly uncertain redshift values.

FIG. 7. -- Camera 2 F160W images of two possible groups among the sample objects. Fields are 19.2″ square, and the coordinates are the J2000 values of the field centers.

TABLE 1

FIELD CENTERS AND SENSITIVITY LIMITS OF FIELDS WITH DETECTED GALAXIES

| α(J2000) | δ(J2000) | Camera | 1 σ Sensitivity, μJy | | | Number of Galaxies |
|---|---|---|---|---|---|---|
| | | | F110W | F160W | F222M | |
| 00 02 24.91 | -00 31 55.5 | 2 | 0.031 | 0.032 | 0.275 | 1 |
| 00 02 25.03 | -00 31 23.0 | 1 | 0.024 | 0.028 | -- | 1 |
| 00 26 15.63 | +10 48 54.3 | 2 | 0.081 | 0.025 | 0.308 | 2 |
| 00 38 05.50 | -02 16 06.0 | 2 | 0.019 | 0.018 | 0.325 | 1 |
| 00 38 17.76 | +48 24 15.1 | 2 | 0.033 | 0.040 | 0.283 | 1 |
| 00 39 52.36 | +48 27 11.9 | 2 | 0.025 | 0.011 | 0.287 | 3 |
| 00 40 46.95 | +41 46 51.3 | 2 | 0.048 | 0.013 | 0.284 | 1 |
| 00 50 09.30 | +32 23 00.2 | 2 | 0.061 | 0.054 | 0.366 | 1 |
| 00 58 10.32 | +30 27 25.7 | 2 | 0.044 | 0.043 | 0.332 | 2 |
| 01 16 34.39 | +33 33 01.2 | 2 | 0.023 | 0.018 | 0.293 | 2 |
| 01 39 11.05 | -17 49 37.6 | 2 | 0.081 | 0.067 | 0.388 | 1 |
| 01 40 09.79 | +01 37 58.1 | 2 | 0.061 | 0.051 | 0.370 | 1 |
| 01 44 14.27 | -15 48 49.4 | 2 | 0.095 | 0.070 | 0.394 | 1 |
| 01 44 26.07 | -15 50 27.4 | 2 | 0.062 | 0.054 | 0.370 | 1 |
| 02 19 49.75 | -02 55 40.5 | 2 | 0.096 | 0.080 | 0.401 | 1 |
| 02 42 48.96 | -34 01 33.8 | 2 | 0.015 | 0.011 | 0.284 | 1 |
| 02 47 57.81 | +19 46 37.8 | 2 | 0.085 | 0.062 | 0.384 | 2 |
| 03 33 20.88 | -09 22 37.9 | 2 | 0.060 | 0.046 | 0.381 | 3 |
| 03 34 13.95 | -36 04 21.5 | 2 | 0.038 | 0.040 | 0.293 | 1 |
| 03 42 46.64 | -29 50 40.1 | 2 | 0.017 | 0.009 | 0.129 | 1 |
| 03 42 46.86 | -29 50 38.0 | 1 | 0.008 | 0.007 | -- | 1 |

| α(J2000) | δ(J2000) | Camera | 1 σ Sensitivity, μJy | | | Number of Galaxies |
|---|---|---|---|---|---|---|
| | | | F110W | F160W | F222M | |
| 04 12 30.26 | -57 36 37.4 | 2 | 0.037 | 0.030 | 0.303 | 1 |
| 04 33 35.49 | +24 04 24.7 | 2 | 0.065 | 0.056 | 0.381 | 1 |
| 05 34 19.82 | -66 24 20.0 | 1 | 0.037 | 0.047 | -- | 1 |
| 05 35 10.06 | +22 03 12.5 | 2 | 0.076 | 0.081 | 0.362 | 1 |
| 05 46 38.38 | -32 21 09.8 | 2 | 0.073 | 0.026 | 0.331 | 1 |
| 06 13 54.05 | +47 37 50.7 | 2 | 0.019 | 0.017 | 0.306 | 1 |
| 06 13 56.33 | +47 37 37.6 | 1 | 0.017 | 0.018 | -- | 1 |
| 06 52 28.66 | -00 13 43.8 | 2 | 0.045 | 0.024 | 0.300 | 1 |
| 06 54 30.91 | +60 44 24.0 | 2 | 0.067 | 0.052 | 0.377 | 2 |
| 07 33 03.20 | -50 26 34.8 | 2 | 0.046 | 0.041 | 0.333 | 1 |
| 07 43 52.58 | +64 17 02.6 | 2 | 0.022 | 0.016 | 0.293 | 2 |
| 07 55 19.89 | +21 52 12.2 | 2 | 0.095 | 0.067 | 0.401 | 1 |
| 07 58 31.05 | -60 42 27.4 | 2 | 0.020 | 0.015 | 0.310 | 1 |
| 08 13 07.30 | +74 56 10.0 | 2 | 0.043 | 0.035 | 0.332 | 1 |
| 08 13 09.43 | +74 57 14.6 | 2 | 0.051 | 0.041 | 0.333 | 2 |
| 08 51 49.63 | +11 52 40.5 | 2 | 0.036 | 0.020 | 0.305 | 1 |
| 08 51 51.90 | +11 52 59.7 | 1 | 0.030 | 0.020 | -- | 1 |
| 10 24 27.07 | +46 59 23.8 | 2 | 0.050 | 0.020 | 0.287 | 3 |
| 11 48 36.96 | +14 29 58.6 | 2 | 0.046 | 0.022 | 0.307 | 1 |
| 11 49 27.58 | +12 47 25.2 | 2 | 0.038 | 0.013 | 0.293 | 4 |
| 12 08 34.93 | -12 24 54.8 | 2 | 0.047 | 0.045 | 0.324 | 1 |
| 12 20 57.18 | +86 17 55.1 | 2 | 0.041 | 0.033 | 0.331 | 1 |
| 12 27 51.39 | +17 02 34.6 | 1 | 0.033 | 0.020 | -- | 1 |

TABLE 1 (CONTINUED)

| α(J2000) | δ(J2000) | Camera | 1 σ Sensitivity, μJy | | | Number of Galaxies |
|---|---|---|---|---|---|---|
| | | | F110W | F160W | F222M | |
| 12 27 58.20 | +17 02 57.4 | 2 | 0.051 | 0.019 | 0.308 | 1 |
| 12 40 24.61 | -41 02 59.4 | 1 | 0.010 | 0.008 | -- | 1 |
| 12 51 36.36 | +56 30 13.9 | 2 | 0.033 | 0.029 | 0.305 | 2 |
| 12 56 23.83 | +22 03 45.1 | 2 | 0.025 | 0.035 | 0.178 | 1 |
| 13 03 48.06 | -25 29 39.4 | 2 | 0.050 | 0.017 | 0.301 | 1 |
| 13 46 55.36 | -11 43 06.4 | 2 | 0.066 | 0.053 | 0.384 | 2 |
| 13 55 35.74 | +18 18 60.0 | 2 | 0.062 | 0.054 | 0.369 | 1 |
| 14 15 55.31 | +52 32 53.5 | 2 | 0.057 | 0.026 | 0.302 | 1 |
| 14 17 15.44 | +52 20 57.7 | 2 | 0.033 | 0.032 | 0.283 | 1 |
| 14 23 15.70 | +38 32 19.3 | 2 | 0.036 | 0.020 | 0.314 | 2 |
| 14 40 36.64 | +53 19 53.2 | 1 | 0.020 | 0.019 | -- | 1 |
| 15 32 15.78 | +32 47 50.4 | 2 | 0.030 | 0.018 | 0.307 | 7 |
| 15 32 23.98 | +32 47 27.5 | 1 | 0.021 | 0.016 | -- | 1 |
| 15 49 26.74 | +21 21 15.2 | 2 | 0.012 | 0.009 | 0.305 | 4 |
| 15 49 27.08 | +21 18 55.0 | 2 | 0.016 | 0.012 | 0.296 | 1 |
| 15 51 16.40 | +32 56 46.6 | 2 | 0.036 | 0.036 | 0.296 | 1 |
| 16 09 38.03 | +65 25 03.3 | 2 | 0.019 | 0.030 | 0.293 | 2 |
| 16 17 31.10 | +32 13 49.3 | 2 | 0.029 | 0.024 | 0.309 | 1 |
| 16 09 38.46 | +65 25 35.4 | 1 | 0.015 | 0.014 | -- | 1 |
| 16 27 34.16 | +82 33 56.5 | 2 | 0.064 | 0.020 | 0.295 | 1 |
| 16 32 05.64 | -13 07 44.4 | 2 | 0.025 | 0.047 | 0.202 | 1 |
| 16 43 20.18 | +17 09 39.5 | 2 | 0.031 | 0.024 | 0.307 | 1 |
| 17 41 47.60 | -53 48 11.0 | 1 | 0.015 | 0.014 | -- | 1 |

TABLE 1 (CONTINUED)

| α(J2000) | δ(J2000) | Camera | 1 σ Sensitivity, μJy | | | Number of Galaxies |
|---|---|---|---|---|---|---|
| | | | F110W | F160W | F222M | |
| 17 57 15.10 | +66 36 53.5 | 2 | 0.061 | 0.053 | 0.370 | 1 |
| 21 16 55.70 | +02 33 30.3 | 2 | 0.058 | 0.050 | 0.332 | 4 |
| 21 43 26.73 | +43 31 54.1 | 2 | 0.063 | 0.024 | 0.326 | 1 |
| 22 09 43.80 | +17 47 28.0 | 2 | 0.092 | 0.068 | 0.392 | 3 |
| 22 17 47.64 | +00 31 45.7 | 2 | 0.028 | 0.032 | 0.277 | 2 |
| 22 30 00.80 | +26 51 26.2 | 2 | 0.032 | 0.030 | 0.309 | 2 |
| 22 50 35.60 | +14 26 20.6 | 2 | 0.059 | 0.046 | 0.329 | 1 |
| 23 47 37.74 | +18 51 10.0 | 2 | 0.083 | 0.065 | 0.379 | 1 |
| Mean Value | | | 0.043 | 0.033 | 0.319 | 1.48 |
| Standard Deviation | | | 0.023 | 0.019 | 0.050 | 1.00 |

TABLE 2

OBJECT SIZES, MAGNITUDES, PHOTOMETRIC REDSHIFTS, AND MORPHOLOGIES[1]

| Object | $r_1$," | V | R | F110W | F160W | F222M | z | Morph. | T |
|---|---|---|---|---|---|---|---|---|---|
| 000224.76-003256.3 | 0.60 | 20.31 | -- | 23.20 | 22.94 | >17.51 | 2.7: | A | |
| 000225.35-003121.7 | 0.36 | 23.39 | -- | 22.14 | 21.36 | -- | 2.1 | S | 1 |
| 002615.22+104856.8 | 0.58 | >20.37 | >20.41 | 21.90 | 21.44 | 20.66 | 1.2 | A | |
| 002616.21+104852.8 | 0.30 | 19.03 | 18.52 | 17.25 | 16.51 | 17.17 | 0.1 | E | -5 |
| 003804.90-021602.8 | 0.42 | -- | -- | 22.73 | 21.26 | 21.49 | 0.9 | S | 2 |
| 003817.70+482432.2 | 0.38 | >22.42 | -- | 23.87 | 24.21 | >18.14 | -- | A | |
| 003951.58+482713.0 | 0.44 | 23.91 | -- | 24.59 | 22.83 | >17.18 | 2.7: | A | |
| 003952.15+482713.0 | 0.26 | 24.04 | 18.55 | 21.26 | 21.06 | 20.40 | 1.2 | S | 5 |
| 003952.23+482717.6 | 0.40 | 22.06 | -- | 23.51 | 24.75 | >17.77 | <0.1: | A | |
| 004046.80+414659.0 | 0.54 | >20.09 | -- | 22.61 | 22.24 | >15.85 | -- | A | |
| 005009.10+322354.5 | 0.26 | 23.84 | -- | 21.32 | 20.96 | 20.87 | 0.7 | A | |
| 005810.36+302729.3 | 0.52 | >21.16 | 23.36 | 22.41 | 22.79 | 21.06 | 0.3 | S | 3 |
| 005810.72+302732.7 | 0.32 | 22.57 | 22.10 | 21.54 | 21.50 | >17.81 | 0.4 | A | |
| 011633.87+333255.3 | 0.57 | 23.66 | 22.34 | 21.83 | 21.14 | 21.37 | 0.6 | A | |
| 011634.53+333305.8 | 0.32 | 24.90 | 23.34 | 22.67 | 22.39 | >16.26 | 0.5 | E | -5 |
| 013910.91-174930.8 | 0.56 | -- | 24.00 | 22.22 | 22.05 | >18.19 | 1.3 | A | |
| 014010.02+013754.8 | 0.42 | >20.50 | -- | 21.89 | 21.22 | 20.78 | 1.4 | S | 2 |
| 014413.98-154850.1 | 0.49 | -- | -- | 21.56 | 21.06 | 20.71 | 1.0: | A | |
| 014426.64-155031.4 | 0.45 | -- | -- | 22.76 | 22.07 | >16.93 | -- | A | |
| 021949.89-025541.2 | 0.48 | 22.27 | -- | 21.07 | 19.85 | 19.89 | 0.3: | E | -5 |
| 024249.29-340131.1 | 0.55 | -- | -- | 23.28 | 23.45 | >15.80 | -- | P | 99 |
| 024757.55+194643.8 | 0.30 | >22.42 | >22.47 | 21.20 | 21.43 | 21.48 | 0.1 | P, D | 99 |
| 024757.82+194626.3 | 0.57 | 23.65 | 22.32 | 21.70 | 20.45 | 20.14 | 0.4 | S | 6 |
| 033320.51-092236.8 | 0.39 | >21.16? | -- | 22.49 | 22.93 | 21.97 | 1.1: | A | |
| 033320.68-092236.3 | 0.39 | >19.38? | -- | 19.68 | 18.97 | 18.84 | 0.7 | E | -5 |

| Object | $r_1$, ″ | V | R | F110W | F160W | F222M | z | Morph. | T |
|---|---|---|---|---|---|---|---|---|---|
| 033321.07-092236.2 | 0.38 | >20.37: | -- | 21.30 | 20.28 | 20.23 | 0.7 | E | -5 |
| 033414.01-360429.9 | 0.42 | -- | -- | 20.86 | 20.06 | 22.09 | 2.4: | E | -5 |
| 034246.44-295035.3 | 0.59 | -- | -- | 21.92 | 22.86 | -- | -- | A | |
| 034247.20-295043.8 | 0.27 | -- | -- | 20.97 | 20.41 | 20.18 | 1.0 | P; M | 99 |
| 041229.42-573633.2 | 0.32 | -- | -- | 21.75 | 21.18 | 21.11 | 0.5 | S | 1 |
| 043335.95+240426.0 | 0.26 | >20.98 | -- | 20.31 | 19.69 | 19.93 | 0.6: | S | 3 |
| 053417.32-662438.7 | 0.23 | -- | -- | 21.44 | 20.95 | -- | -- | A | |
| 053510.72+220308.2 | 0.53 | 18.32 | -- | 20.85 | 20.00 | 19.74 | 0.9 | S | 5 |
| 054639.01-322120.4 | 0.50 | -- | -- | >22.61 | 21.57 | 22.02 | -- | S | 3 |
| 061353.30+473750.5 | 0.49 | >20.94 | >20.99 | 23.45 | 23.41 | >16.67 | -- | A | |
| 061356.41+473739.6 | 0.18 | >22.04 | >22.09 | 22.18 | 23.33 | -- | -- | A | |
| 065228.62-001338.0 | 0.50 | >21.13 | >21.18 | >21.81 | 21.24 | 21.12 | -- | A; D | |
| 065430.81+604417.0 | 0.47 | >21.70 | 23.90 | 22.09 | 22.82 | >17.45 | 0.2: | A | |
| 065432.08+604421.0 | 0.45 | >20.98 | 22.14 | 21.75 | 20.96 | 20.96 | 0.6 | P | 99 |
| 073303.15-502636.3 | 0.53 | -- | -- | 22.66 | 21.93 | 21.39 | 1.0: | A | |
| 074351.72+651658.6 | 0.34 | -- | -- | 19.46 | 18.79 | 18.95 | 0.5 | S | 5 |
| 074352.28+651707.8 | 0.56 | -- | -- | >21.17 | 22.29 | >15.76 | -- | S | 4 |
| 075519.67+215212.8 | 0.50 | >21.70 | 23.47 | 21.92 | 21.79 | 21.64 | 1.0: | A | |
| 075831.38-604240.6 | 0.54 | -- | -- | 23.75 | 23.23 | >17.19 | -- | A | |
| 081305.30+745601.0 | 0.40 | -- | -- | 20.94 | 20.32 | 19.82 | 1.0 | E | -5 |
| 081309.25+745719.4 | 0.55 | -- | -- | 21.32 | 21.06 | 20.86 | 1.0 | A | |
| 081311.25+745707.8 | 0.31 | -- | -- | 18.77 | 17.94 | 17.98 | 0.6 | S | 3 |
| 085149.69+115234.6 | 0.48 | >21.43: | >21.48: | 23.56 | 22.68 | >17.19 | -- | A | |
| 085152.05+115228.0 | 0.32 | >22.04: | >22.09: | 22.18 | 23.33 | -- | -- | A | |

| Object | $r_1$," | V | R | F110W | F160W | F222M | z | Morph. | T |
|---|---|---|---|---|---|---|---|---|---|
| 102427.40+465933.2 | 0.63 | >20.55 | >20.60 | >22.15 | 22.39 | >16.32 | -- | S | 7 |
| 102427.45+465927.9 | 0.50 | >22.21 | >22.25 | 23.91 | 23.23 | >17.98 | -- | S | 3 |
| 102427.72+465921.2 | 0.48 | 23.31 | 23.54 | 22.92 | 23.45 | >17.83 | 0.2: | A | |
| 114836.98+142951.6 | 0.32 | -- | -- | >22.60 | 22.56 | >17.19 | -- | A | |
| 114927.72+124728.4 | 0.60 | -- | -- | >21.73 | 22.82 | >16.32 | -- | A | |
| 114927.82+124716.7 | 0.42 | -- | -- | >22.61 | 23.76 | >17.19 | -- | A | |
| 114927.86+124718.3 | 0.40 | -- | -- | >22.61 | 22.95 | >17.19 | -- | A | |
| 114929.42+124716.5 | 0.63: | -- | -- | >21.90 | 23.08 | >16.49 | -- | A | |
| 120834.64-122455.3 | 0.35 | -- | -- | 19.85 | 18.93 | 18.73 | 0.8: | S | 0 |
| 122101.96+861758.0 | 0.31 | -- | -- | 22.15 | 21.17 | 21.77 | 0.8 | S | 2 |
| 122751.33+170229.0 | 0.21 | 22.46 | -- | 21.18 | 20.57 | -- | 0.6 | S | 2 |
| 122758.60+170250.6 | 0.43 | 20.43 | -- | 19.25 | 19.04 | >13.52 | <0.1 | I | 10 |
| 124024.80-410228.9 | 0.32 | -- | -- | 21.89 | 21.21 | -- | -- | S | 1 |
| 125135.51+563015.5 | 0.26 | -- | -- | 20.74 | 20.10 | 20.07 | 0.5: | S | 1 |
| 125137.13+563016.4 | 0.64 | -- | -- | 21.58 | 22.14 | >15.99 | -- | I | 10 |
| 125624.12+220452.7 | 0.51 | >18.95 | -- | 21.66 | 20.82 | 21.46 | 0.6: | S | 4 |
| 130348.26-252928.3 | 0.54 | -- | -- | >22.02 | 22.38 | >16.67 | -- | A | |
| 134655.14-114311.6 | 0.26 | -- | -- | 21.61 | 20.77 | 20.29 | 1.0 | A | |
| 134655.77-114303.4 | 0.83 | -- | -- | 21.93 | 20.61 | 20.59 | 0.9 | S | 4 |
| 135535.50+181905.9 | 0.46 | -- | -- | 21.34 | 20.72 | 21.11 | 0.6 | S | 3 |
| 141556.01+523254.7 | 0.52 | -- | -- | >21.74 | 22.17 | >16.32 | -- | A | |
| 141715.05+522050.1 | 0.34 | -- | -- | 19.13 | 18.53 | 18.55 | 0.5 | E | -5 |
| 142315.18+383225.9 | 0.31 | -- | -- | 20.99 | 20.47 | 20.66 | 0.6 | S | 0 |
| 142316.32+383225.4 | 0.33 | -- | -- | 19.13 | 18.29 | 18.31 | 0.6 | E | -5 |
| 144036.60+532020.9 | 0.27 | -- | -- | 22.01 | 21.53 | -- | -- | S | 4 |
| 153215.08+324750.2 | 0.40 | >21.43 | >21.48 | 23.38 | 23.50 | >17.20 | -- | A | |

| Object | $r_1$, ″ | V | R | F110W | F160W | F222M | z | Morph. | T |
|---|---|---|---|---|---|---|---|---|---|
| 153215.54+324753.7 | 0.56 | >22.42 | >22.47 | >24.01 | 24.05 | >18.19 | -- | A | |
| 153214.98+324750.0 | 0.51 | >21.43 | >21.48 | 23.85 | 23.45 | >17.20 | -- | A | |
| 153215.95+324743.8 | 0.26 | 22.32 | 21.46 | 20.48 | 19.45 | 19.75 | 0.6 | S | 2 |
| 153216.11+324747.7 | 0.44 | 23.29 | 23.23 | >23.02 | 23.32 | >17.20 | 1.8: | A | |
| 153216.24+324749.7 | 0.57 | >22.94 | >22.98 | >24.53 | 24.06 | >18.70 | -- | A | |
| 153216.45+324747.2 | 0.40 | 24.30 | 23.92 | >23.65 | 23.05 | >17.82 | 0.4: | A | |
| 153223.82+324724.1 | 0.53 | >19.44 | 23.82 | 23.70 | 21.84 | -- | 2.7: | A | |
| 154926.26+212113.9 | 0.47 | -- | -- | >21.09 | 21.60 | >15.69 | -- | S | 3 |
| 154926.34+212112.3 | 0.46 | -- | -- | 23.39 | 23.11 | >17.46 | -- | A | |
| 154926.78+212116.8 | 0.42 | -- | -- | 22.02 | 21.36 | 22.07 | 0.6: | S | 2 |
| 154927.04+212111.3 | 0.43 | -- | -- | 23.56 | 22.70 | 22.05 | 1.1 | A | |
| 154927.20+211848.2 | 0.25 | -- | -- | 21.25 | 20.62 | 21.20 | 0.6 | P; M | 99 |
| 155116.54+325655.9 | 0.54 | -- | -- | >21.36 | 21.88 | >15.94 | -- | A | |
| 160936.82+652500.3 | 0.30 | -- | -- | 20.53 | 20.03 | 20.26 | 0.6 | S | 2 |
| 160936.92+652506.9 | 0.41 | -- | -- | 21.09 | 20.17 | 20.48 | 0.6 | S | 3 |
| 160937.81+652456.5 | 0.34 | -- | -- | 22.00 | 21.61 | -- | -- | S | 8 |
| 161731.28+321354.5 | 0.35 | -- | -- | 20.79 | 19.97 | 20.59 | 0.6 | S | 3 |
| 162734.65+823405.8 | 0.47 | -- | -- | >21.68 | 21.87 | >16.30 | -- | P | 99 |
| 163205.92-130736.3 | 0.48 | -- | -- | 21.27 | 20.10 | 20.19 | 0.8 | P; D | 99 |
| 164319.99+170939.6 | 0.59 | -- | -- | 22.33 | 21.52 | 21.25 | 0.9 | S | 7 |
| 174145.12-534804.9 | 0.24 | -- | -- | 21.18 | 20.51 | -- | -- | S | 4 |
| 175716.94+663648.8 | 0.58 | -- | -- | 20.77 | 20.85 | >16.77 | -- | A | |
| 211655.38+023333.2 | 0.54 | >22.06 | >22.10 | 22.63 | 21.51 | 20.86 | 1.5 | E | -5 |
| 211655.39+023332.0 | 0.54 | >22.06 | >22.10 | 22.57 | 21.07 | 20.43 | 1.6 | A | |
| 211655.43+023329.5 | 0.52 | 21.29 | 21.34 | 21.72 | 21.74 | 21.25 | 1.1: | A | |
| 211655.54+023326.8 | 0.53 | 22.43 | 23.64 | 21.15 | 20.17 | 19.78 | 0.9 | S | 4 |

| Object | $r_1$, ″ | V | R | F110W | F160W | F222M | z | Morph. | T |
|---|---|---|---|---|---|---|---|---|---|
| 214327.23+433201.2 | 0.41 | -- | -- | 19.87 | 19.23 | 19.40 | 0.7 | E | -5 |
| 220943.69+174720.7 | 0.52 | >22.06 | >22.10 | 22.01 | 21.21 | >17.81 | -- | A | |
| 220944.14+174716.1 | 0.55 | >20.17 | >20.22 | 22.57 | 21.07 | 20.43 | 1.0 | P | 99 |
| 220944.35+174718.1 | 0.37 | >22.06 | >22.10 | 21.31 | 21.01 | 21.57 | 1.1: | A | |
| 221748.12+003138.1 | 0.49 | 22.94 | -- | 23.20 | 23.19 | 18.69 | ? | A | |
| 221747.67+003142.3 | 0.46 | >20.46 | -- | 21.37 | 20.65 | 20.33 | 0.9: | A | |
| 223000.47+265121.3 | 0.48 | -- | -- | 22.36 | 21.55 | >17.19 | -- | A | |
| 223000.78+265128.5 | 0.41 | 22.31 | -- | 20.39 | 19.90 | 19.84 | 0.3 | E | -5 |
| 225035.32+142620.5 | 0.43 | 21.70 | -- | 20.24 | 19.44 | 19.16 | 0.3 | S | 4 |
| 234737.56+185106.3 | 0.52 | 23.78 | -- | 21.81 | 20.85 | 20.98 | 0.4 | S | 3 |

[1] Morphologies are coded as follows: A = Ambiguous, D = Double, E = Elliptical, I = Irregular, M = Merger, P = Peculiar, S = Spiral. Maximum errors are ~ ± 0.1″ in $r_1$, ~ ± 0.1 in magnitudes, ~ ± 0.3 in redshift. Objects with no magnitude for F222M were observed with Camera 1.

TABLE 3

AXIAL RATIOS AND MAJOR AXIS POSITION ANGLES OF SPIRAL GALAXIES[1]

___________________________________________________________________________

| Object | a/b | P.A.,° |
|---|---|---|
| 000225.35-003121.7 | 1.0 | -- |
| 003804.90-021602.8 | 2.4 | 60 |
| 003952.15+482713.0 | 2.5 | 170 |
| 005009.10+322354.5 | 2.2 | 0 |
| 005810.36+302729.3 | 1.0 | -- |
| 014010.02+013754.8 | 1.2 | 80 |
| 024757.82+194626.3 | 2.9 | 160 |
| 041229.42-573633.2 | 1.0 | -- |
| 043335.95+240426.0 | 1.5 | 160 |
| 053417.32-662438.7 | 1.6 | 150 |
| 053510.72+220308.2 | 3.4 | 70 |
| 054639.01-322120.4 | 1.1 | 160 |
| 074352.28+651707.8 | 3.1 | 20 |
| 074351.72+651658.6 | 1.6 | 50 |
| 081311.25+745707.8 | 1.6 | 20 |
| 102427.40+4659.332 | 1.2 | 130 |
| 102427.45+465927.9 | 1.1 | 50 |
| 120834.64-122455.3 | 2.1 | 110 |
| 122101.96+861758.0 | 2.8 | 40 |
| 122751.33+170229.0 | 1.2 | 30 |
| 124024.80-410228.9 | 1.4 | 30 |
| 125135.51+563015.5 | 1.5 | 130 |
| 125624.12+220452.7 | 3.3 | 140 |
| 134655.77-114303.4 | 1.5 | 130 |

TABLE 3 (CONTINUED)

| Object | a/b | P.A.,° |
|---|---|---|
| 135535.50+181905.9 | 2.4 | 10 |
| 142315.18+383225.9 | 1.3 | 90 |
| 144036.60+532020.9 | 1.3 | 70 |
| 153215.54+324743.8 | 1.7 | 140 |
| 154926.26+212113.9 | 1.0 | -- |
| 154926.78+212116.8 | 2.2 | 110 |
| 160936.82+652500.3 | 2.5 | 10 |
| 160936.92+652506.9 | 2.7 | 30 |
| 160937.81+652456.5 | 1.9 | 90 |
| 161731.28+321354.5 | 2.2 | 120 |
| 163205.92-130736.3 | 1.1 | 40 |
| 164319.99+170939.6 | 3.7 | 50 |
| 174145.12-534804.9 | 2.0 | 20 |
| 211655.54+023326.8 | 2.4 | 150 |
| 225035.32+142620.5 | 1.2 | 90 |
| 234737.56+185106.3 | 1.2 | 150 |

[1] Errors range from ~ ± 0.1 - 0.3 in axial ratio and ~ ± 10° - 30° in P.A.

TABLE 4

CANDIDATE GALAXY PAIRS

___________________________________________________________________________

| Objects | z | Morph. | Separation, ″ | Separation, kpc [1] |
|---|---|---|---|---|
| 005810.36+302729.3 | 0.3 | S | | |
| | | | 5.8 | 34 |
| 005810.72+302732.7 | 0.4 | A | | |
| 011633.87+333255.3 | 0.6 | A | | |
| | | | 13.4 | 105 |
| 011634.53+333305.8 | 0.5 | E | | |
| 033320.68-092236.3 | 0.7 | E | | |
| | | | 5.8 | 53 |
| 033321.07-092236.2 | 0.7 | E | | |
| 134655.14-114311.6 | 1.0 | A | | |
| | | | 12.4 | 123 |
| 134655.77-114303.4 | 0.9 | S | | |
| 142315.18+383225.9 | 0.6 | S | | |
| | | | 13.4 | 115 |
| 142316.32+383225.4 | 0.6 | E | | |

| Objects | z | Morph. | Separation, ″ | Separation, kpc |
|---|---|---|---|---|
| 160936.82+652500.3 | 0.6 | S | | |
| | | | 6.6 | 56 |
| 160936.92+652506.9 | 0.6 | S | | |
| 211655.38+023333.2 | 1.5 | E | | |
| | | | 1.2 | 13 |
| 211655.39+023332.0 | 1.6 | A | | |
| 220944.14+174716.1 | 1.0 | P | | |
| | | | 3.6 | 37 |
| 220944.35+174718.1 | 1.1: | A | | |

[1] Measured using $H_0 = 50$ km s$^{-1}$ Mpc$^{-1}$ and $q_0 = 0.1$, and assuming both objects to be at the lower redshift of the pair if the redshifts are unequal.

137